\documentclass[aps,prl,twocolumn]{revtex4}
\usepackage{psfig}
\newtheorem{theorem}{Theorem}
\newtheorem{proposition}{Proposition}

\def\duzomniejsze{<\kern-.7mm<}
\def\duzowieksze{>\kern-.7mm>}

\def\textbf#1{{\bf #1}}
\def\beq{\begin{equation}}
\def\eeq{\end{equation}}
\def\be{\begin{equation}}
\def\ee{\end{equation}}
\def\ben{\begin{eqnarray}}
\def\een{\end{eqnarray}}
\def\beqa{\begin{eqnarray}}
\def\eeqa{\end{eqnarray}}
\def\eea{\end{array}}
\def\bea{\begin{array}}
\newcommand{\bei}{\begin{itemize}}
\newcommand{\eei}{\end{itemize}}
\newcommand{\bee}{\begin{enumerate}}
\newcommand{\eee}{\end{enumerate}}

\def\n{{\otimes n}}

\def\<{\langle}
\def\>{\rangle}

\def\dt#1{{{\kern -.0mm\rm d}}#1\,}
\def\ypodpis{\raise4mm\hbox{$\omega$}}

\def\ourset{HS }
\def\ourmaps{Hyper-Maps }
\def\mapsfree{LOCC+\ourset }
\def\efppt{E^{ppt}_f }

\begin{document}
\title{Are the laws of entanglement theory thermodynamical?}
\author{Micha\l{} Horodecki$^{(1)}$, Jonathan Oppenheim$^{(1)(2)}$,
and Ryszard Horodecki$^{(1)}$}

\affiliation{$^{(1)}$Institute of Theoretical Physics and Astrophysics,
University of Gda\'nsk, Poland}
\affiliation{$^{(2)}$
Racah Institute of Theoretical Physics, Hebrew University of Jerusalem, Givat Ram, Jerusalem 91904, Israel}


\begin{abstract}
We argue that on its face, entanglement theory satisfies laws equivalent to thermodynamics
if the theory can be made reversible by adding certain {\it bound entangled
states} as a free resource during entanglement manipulation.  Subject to plausible assumptions,
we prove that this is not the case in general, and discuss the implications of this for the
{\it thermodynamics of entanglement}.
\end{abstract}
\maketitle


The pioneering papers in quantum information theory
\cite{BBPSSW1996,BDSW1996,BBPS1996} revealed a potential
irreversibility in entanglement processing. They suggested
that  by local operations  and classical communication (LOCC),
one needs more pure entangled states to produce a state
than can be drawn from it. Some researchers expressed the intuition
that this is due to some "second law" in entanglement processing.
The first paper making this analogy rigorous was
\cite{popescu-rohrlich}. However the authors did not discuss
irreversibility, instead building the analogy in the region of
full reversibility - in the domain of pure states
(see also \cite{PlenioVedral1998}). It was sometimes argued
that irreversibility is where entanglement and
thermodynamics differ, as e.g. the Carnot cycle is reversible.
In \cite{termo} a different point of view was presented
which attempted to account for this irreversibility.
The leading idea was that entanglement is analogous to {\it energy} and
that distillation of pure entanglement is like drawing work.
(the amount of pure entanglement drawn from a state is $E_d$
while the amount needed to create a state we denote by $E_c$).
Meanwhile the extreme case of irreversibility was discovered:
the bound entangled (BE) state. One cannot draw
any pure entanglement from them, but entanglement is needed to create them.

In this paper we will investigate aspects of the latter analogy and argue that it is extremely useful
for understanding the basic laws of entanglement processing. We
study the consequences of our proposal, and make the theory precise so that testing is possible.
We first show that the very fact of irreversibility
does not cancel the analogy, rather it is a constitutive element.  This enables
us to state the three laws of entanglement for such a theory.
Next we however show that the analogy with reversible thermodynamics
does not appear hold in general.  We then speculate on possible ways the
basic laws of thermodynamics may still yield insights into quantum information
theory.

\begin{figure}
\psfig{file=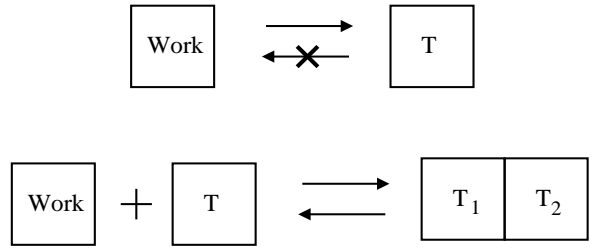,width=8cm}
\caption[l]{Reversible and irreversible processes
in thermodynamics}
\label{fig:rev-irrev}
\end{figure}

The main observation we will need is that what constitutes
thermodynamics,  and distinguishes it e.g. from  mechanics, is that there
are {\it two  forms} of energy: disordered (heat) and ordered
(work). The first law does not ``feel'' any difference
between these. It says that to create a single heat bath of
energy $E$ we need precisely this amount of work. The second law
says that this creation is {\it irreversible}: from a single heat
bath one cannot draw work.  Thus the second law accounts for the basic
irreversibility of the heat bath formation: it is due to loss
of information.
In its ideal version, thermodynamics offers also
a reversible change of work into heat and vice versa (the Carnot cycle).
However
this is only possible  if one has heat reservoirs from the very beginning.
This is depicted on Fig. \ref{fig:rev-irrev}.
To summarize: 1) there is a form of energy, that cannot be used
to draw work (single heat baths); 2) it can however be used to store work, but: 3) work can be stored
in heat only if there is some  heat at the beginning 4) in the
latter case work can  be stored {\it reversibly}.

Let us now argue that thus far,
entanglement processing not only does not exclude
a perfect analogy with the features of thermodynamics outlined above,
but suggests strongly that there is one. (We will then provide
a new result, which, up to a plausible conjecture, precludes
such a prima-facia possibility.)  We will follow an idea related to the entanglement-energy
analogy of \cite{termo}.

First it is reasonable to assume that pure state entanglement
is analogous to mechanical energy,
while mixed state entanglement corresponds to energy that may be partly accumulated in
the form of heat.
Now, distillation \cite{BBPSSW1996}
i.e.   drawing pure entanglement from
mixed states  is like running an engine that produces
work out of heat.
There is a question: in thermodynamics  we have systems  that
{\it have} energy but  no work can be drawn from them - single heat
baths. If we didn't have
a counterpart of it in entanglement processing, then a  constitutive
element would be missing.
However  there exist
so-called {\it bound entangled} states
\cite{bound} that are entangled (have ``energy'')
but this ``energy'' cannot be used i.e. changed into an ordered form.
Thus BE states are the counterpart of  a ``single heat bath'' - a completely
disordered form of entanglement.  Other mixed states would
represent partially ordered entanglement: work stored in thermodynamical
systems
(e.g.  pairs of reservoirs of different temperatures).
The process of forming a BE state out of pure states could
be thought of as analogues to Joule's experiment establishing the equivalence
between work and heat where  all work is (irreversibly) changed into
heat. The energy of the created heat bath is equal to
the amount of work added to the system. This leads us to the conclusion:
{\it The counterpart of total energy would be entanglement cost (amount of
pure entanglement needed to produce the state).}  At this moment one can
ask a pointed question: does a counterpart to the first law hold in
entanglement theory?
Is the
entanglement in a BE state still present? Or perhaps
to form the state one needs entanglement but during the formation
process it gets dissipated \cite{first-law}. 
This is very much connected with the main question which we
will now address:
{\it Can we have an analogue of reversible work extraction?} One
might say: `no,
because in entanglement processing there is irreversibility'.
This is however not an adequate answer.  Let us argue
that we can have a perfect analogy of reversible work extraction,
despite the fact that generically we need more singlets to
create a state than can be draw from it.

First note, that any thermodynamical system has more energy than we
can draw from it (excluding the case when one of the reservoirs
has zero temperature). This is due to the second law. The same is true for
entanglement:  for a generic mixed state $E_D<E_c$.  Thus
the {\it irreversibility} of entanglement processing could be due
to a second law of entanglement that says ``the disorder of entanglement
can only increase''. Now, according to Figure \ref{fig:rev-irrev},
for a given state $\varrho$, two conditions should be matched
\bei
\item one should be able to distill $E_d$
of pure entanglement, but in such a way that the garbage left over from
the distillation procedure
(denoted by $\varrho_g$)
is a BE state with $E_c(\varrho_g)=E_c(\varrho) - E_d(\varrho)\equiv
E_b(\varrho)$
\be
\varrho \to \varrho_g \otimes \psi_+^\n.
\ee
\item one should be able to form the state $\varrho$
out of $E_d(\varrho)$
singlets plus a BE state of  entanglement cost equal to $E_b(\varrho)$
\eei

This is a consequence of {\it conservation of $E_c$} which would be analogous to the {\it first law}.
Even though we will show that in general this will not hold,
we think it is important to realize how this theory would
look. Indeed perhaps reversible thermodynamics of entanglement
is an important element of the total, more complicated picture.
Besides, the independent development
\cite{AdenauertPE-PPT} suggests that  for some classes
of mixed states this theory may be valid.

Thus, we would have the following  reversible
thermodynamics of entanglement:
\bei
\item There is a unique measure of entanglement ($E_c$)
\item $E_d$ is not an autonomous measure of entanglement:
it is the ``work'' that can be drawn in units of $E_c$.
\item Processes of formation and distillation are {\it reversible} in the sense
described above.
\item The process of changing pure entanglement into mixed entanglement is
{\it irreversible} \cite{locc-orth}.
\eei
Exactly as in statistical physics we would have two kinds of irreversible
processes: dissipation (where $E_c$ decreases)
and pure decoherence (where $E_c$ is constant, but $E_d$ decreases).
The first irreversibility  does not exist in optimal
processes (which we consider).  The second could
be removed if the processes start with some
initial supply of BE states. Then to form a state, one would not
need to create disordered entanglement out of the ordered form
but only {\it dilute} the ordered one into the noisy entanglement
of BE states.

Usually hypothetical reversibility in entanglement processing
is associated with a unique measure of entanglement, in the sense
that there is only one function monotonic under operations.
Indeed we can state the above in such a way that there will only be
one measure of entanglement. Since bound entanglement
is used only as  a source of fully disordered entanglement, it is
not actually a  resource. Thus we can treat it as
free of charge (say heat is cheap, only work costs).
Our class of operations would then be as
postulated in \cite{activation}: LOCC plus BE ancillas.
This does not make the theory
trivial, since we still cannot obtain singlets for free.
Then however, the only monotone would be $E_D$.
Actually, it would be equal to the (regularised) relative entropy distance to
the set of nondistillable states
\cite{Vedral-distbound,Rains1999,Rains-erratum1999}. At this point we
can even formulate a precise theorem
\begin{theorem}
\label{th:relative}
If under a given class of operations $C$ that includes mixing states,
one can reversibly transform $\varrho$
into pure states, then
any asymptotically continuous function $E$
that is monotonic under the class $C$  satisfies
\be
E^\infty(\varrho)=E^\infty_R(\varrho)
\ee
where $E_R^C$ is the relative entropy distance from the set of
states that is closed under $C$.

\end{theorem}
The above notation indicates the regularization of a function $M(\varrho)$
given by $M^\infty(\varrho)=lim_{n\to \infty}{1\over n}M(\varrho^\n)$

{\bf Remark.} For such states we thus have a unique
entanglement measure equal
to the familiar regularized relative entropy of entanglement.

{\bf Proof.} Uniqueness comes from reversibility and uniqueness
of the entanglement measure in the pure state case
\cite{popescu-rohrlich,DonaldHR2001} (that again comes from reversibility
between pure states \cite{BBPS1996}).
$E_R^C$ is a  monotone under $C$\cite{Michal2001}, and is also asymptotically continuous
if the considered set is convex\cite{DonaldHR2001} (which is the case since $C$ includes mixing).

Thus we can formulate the {\it second law}.
Surprisingly it has the same form
as in statistical physics: it is Uhlmann monotonicity see e.g.
\cite{Alicki-Lendi} saying that {\it under completely positive maps
the relative entropy does not increase}. Indeed
from this, one gets that the relative entropy distance is monotonic,
from which our theorem follows. In this way the law gives
the optimal entanglement that
can be distilled exactly as the second law gives
the Carnot efficiency. This also suggests a feedback
to thermodynamics: it seems possible to
express the extractable work by means
of the relative entropy distance from the single heat bath states
(for a fixed Hamiltonian).
Indeed, theorem \ref{th:relative}
appears to be satisfied in thermodynamics.
We hope to come back to this interesting
point elsewhere.  It is also rather amusing that in this theory
we would have an analogue to the {\it third law} of thermodynamics:
{\it one cannot distill singlets with perfect fidelity}

Let us return to the question of whether we can have reversibility
for all states.  Recall that for this proposal: {\it Reversible thermodynamics of
entanglement = possibility of reversible separation of
bound and pure entanglement}.

Below we show a counterexample .
To this end we need to be more rigorous. The first problem
is that if there are NPT \cite{NPT} bound entangled states, then
BE cannot be for free, as such states together with some PPT states are
distillable, so that singlets would be for free too. Thus
we shouldn't speak about BE states, but rather about the
smaller class of states that is closed under LOCC {\it and}
under tensor product.
Call this set the Hyper-Set (\ourset). Thus to produce a counter-example,
we should show that there is
a state for which $E_d^{\mapsfree} < E_c^{\mapsfree}$,
where \mapsfree is LOCC operations plus states from \ourset
as free ancillas.  One can  also consider the class of \ourmaps that
includes the class \mapsfree - namely the maps that
do not move states
outside of \ourset or the subclass of these consisting of the
maps that are closed under tensor product.

We do not know the set \ourset or the class of Hyper-Maps.
However, the following result\cite{wolf}
adds an additional restriction: any NPT state
can be distilled with the help of some PPT state.
The \ourset would therefore not be able to include PPT states
and any NPT state.  The conditions on other sets are also
very restrictive, and we therefore conjecture that the \ourset is PPT 
and the class is
PPT maps \cite{PPTops} introduced in \cite{Rains2001}.

Let us therefore define PPT entanglement of formation $E^{ppt}_f$.
Recall that the traditional entanglement of formation was defined as
\cite{BDSW1996}
$E_f(\varrho)=\inf\sum_ip_iS(\varrho^i_{A})$
where the infimum runs over all ensembles
$\varrho=\sum_ip_i|\psi_i\>\<\psi_i|$  and $\varrho_{A}^i$
is obtained by partial trace of $\psi_i$. One knows that the entanglement
cost
is given by the regularized entanglement of formation\cite{cost}
\beq
E_c=E_f^\infty.
\label{eq:cost}
\eeq
To define the PPT version of $E_f$ we need the notion of
{\it phase-separated states}.
A state is called phase-separated if it is either pure or PPT
or a tensor product of both. Then we define
\be
E^{ppt}_f(\varrho)=\inf \sum_ip_i f(\varrho_i)
\ee
where the infimum runs over decompositions of $\varrho$
into phase-separated states, and $f(\varrho^{ppt}_{AB})=0$,
$f(\psi_{AB})=S(Tr_A(|\psi\>_{AB}\<\psi|)=
f(|\psi\>_{AB}\<\psi|\otimes \varrho^{ppt}_{A'B'})$, i.e.
we count only the pure entanglement.  This quantity intuitively
measures the entanglement cost under LOCC+PPT, like entanglement
of formation $E_f$ 
was associated with entanglement
cost under LOCC.
Indeed following \cite{cost}
\begin{proposition}
If $E_f^{ppt}$  is monotonic under PPT maps, and asymptotically
continuous \cite{cont},
then its regularization
is  equal to the entanglement cost
under PPT maps.
\end{proposition}
If the assumptions hold, then we have counterpart of (\ref{eq:cost})
for PPT maps. Consider now the state
\be
\varrho= p|\psi_+\>\<\psi_+|+(1-p)|\psi_-\>\<\psi_-|,
\quad p\in[0,1]
\label{state}
\ee
with $\psi_\pm={1\over \sqrt2}(|00\>\pm|11\>)$. We have
$E_f= E_c=H({1\over 2} + \sqrt{p(1-p)})$ \cite{Vidal-cost2002} where
$H(x)=-x\log x - (1-x) \log (1-x)$ is the binary entropy.
Also $E_d^{ppt}=E_R^{ppt}=S(\varrho_A) - S(\varrho)=1-H(p)$
\cite{BDSW1996,Rains1999}.

We will now state the main result of the paper
\begin{theorem}
For the state (\ref{state}) $p\not=1/2,0,1$ the PPT distillable  
entanglement  is strictly smaller  than the regularized PPT entanglement 
of formation
\end {theorem}
 
{\bf Proof.} We will prove that $\efppt={\efppt}^\infty=E_c$.
We will actually show that it  is true for all  maximally 
correlated (MC) states  \cite{Rains1999} defined as $\varrho=\sum_{ij}a_{ij} |ii\>\<jj|$.
We need the following facts about MC states:
(1) A state $\varrho_{AB}\otimes \sigma_{A'B'}$ is MC iff $\varrho$
and $\sigma$  are MC; (2) any state in the support of an MC state is MC;
(3) if a MC state is PPT then it is separable.
To see (1) note that (i) all pure states in the support of MC state
have the same Schmidt decomposition (SD); (ii) a state is MC if its
eigenvectors have the same Schmidt decomposition.
Thus if $\varrho_{AB}\otimes \sigma_{A'B'}$ is MC, then its eigenvectors
$\psi_{AB}^i\otimes \phi_{A'B'}^i$  (where $\psi_{AB}^i$ and
$\phi_{A'B'}^i$ are eigenvectors of  $\varrho_{AB}$ and $\sigma_{A'B'}$
resp.) have the same SD. This is possible only if
it is true for vectors $\psi_{AB}^i$ and $\phi_{A'B'}^i$ separately. Thus
both $\varrho_{AB}$ and $\sigma_{A'B'}$ have to be MC. Similarly
the converse holds. Fact (2) can be found e.g. in
\cite{Rains1999}. Proof of (3) is straightforward.

Now we can prove the theorem. From  (1) it follows that 
if $\varrho$ is MC then so is $\varrho^\n$.
Now take any ensemble $\varrho=\sum_ip_i \varrho_i$, with 
phase-separated $\varrho_i$. Due to (2) we have that 
all $\varrho_i$ are MC. Then (3) implies that if $\varrho_i$ 
is PPT, it must be separable. Otherwise it is either  
pure or product $|\psi\>\<\psi|\otimes \sigma_{ppt}$
of a pure state and a PPT state. However due to (1), $\sigma_{ppt}$
must be MC, hence it is separable. Thus we do not have PPT states
in decomposition, so that $\efppt(\varrho^\n)=E_f(\varrho^\n)$
hence ${\efppt}^\infty=E_f^\infty(\varrho)$. However we have
\cite{Vidal-cost2002,MichalSS2002} $E_f^\infty=E_f$.
Of course, for states (\ref{state}) with $p\not=0,1/2,1$ we have
$E_d^{ptt}<E_c$. This ends the proof.

Thus, under the conjectures that $\efppt$  is monotonic under PPT and
asymptotically continuous and that \ourset is PPT,
we obtain that there is no
reversible thermodynamics of entanglement. Instead we in general
have some dissipation of entanglement even in optimal processes.
Since $E_R^\infty$ is the entanglement that could be  distilled  in
the reversible case one can be tempted to argue that the difference between
$E_c$ and $E_R^\infty$ gives the amount of entanglement
dissipated during formation, while $E_R^\infty-E_d$
is that dissipated during distillation. Then the process of formation
of BE states could be non-dissipative, while
for maximally  correlated states, the whole bound entanglement would be
dissipated during formation of the state. I.e. bound entanglement
is needed to create the state, but is lost during creation (this is
supported by the fact that one can localize the information
corresponding to the bound entanglement for these state \cite{OHHH2001}).

Does it mean that thermodynamical analogies should be abandoned?
First of all, let us emphasize that even if we do not
have reversible thermodynamics of entanglement
it is definitely instructive to know how
far we are from this regime. Thus one might be able to understand
the theory looking at deviations from the desired behavior.
Moreover, there is still place for the analogy. 
It may be that one can account for the $E_c$ that is dissipated.
Perhaps we sometimes operate in
the non-equilibrium regime, or perhaps when distilling entanglement
we have phase transitions (cf. \cite{OHHphase}).  
Then we may have irreversiblity
due to the release of heat.
A result disproving the conjecture that $\efppt$ is asymptotically continuous
would also reopen the possibility that these three laws may
still hold as is.

Finally,  in a recent development\cite{AdenauertPE-PPT} a nontrivial state
was exhibited, for  which  there is reversibility under PPT maps,
even though $E_d<E_c$. If one can show that
the PPT map can be realized by means of LOCC+PPT, we would obtain a nontrivial regime
where reversible thermodynamics of entanglement holds.

To summarize, the features that distinguish entanglement theory
from reversible thermodynamics are not due to the existence of
bound entanglement (or distillation-formation irreversibility).
Rather the problem is that two ``phases'', bound entanglement and
free (pure) entanglement cannot in general be {\it separated}
without loss of entanglement cost. Finally, we believe that our paper
provides a new, clearer   picture of entanglement theory, and that
further investigation into the thermodynamics of entanglement is warranted.

{\bf Acknowledgments}:
This work is supported by EU grant EQUIP, No. IST-1999-11053.
J.O. acknowledges the support of
Lady Davis, and
ISF grant 129/00-1.  We thank K. and P. Horodecki,
and A. and U. Sen for discussions.  J.O. also thanks J. Bekenstein and J. Eisert.


\begin{thebibliography}{29}
\bibitem[{loc()}]{locc-orth}
\bibinfo{note}{As noted in \cite{termo} this does not concern mixed states that
  are locally orthogonal mixures of pure states. We hope to elaborate on this
  it elswhere.}
\bibitem[{NPT()}]{NPT}
\bibinfo{note}{A state is PPT (NPT) if its partial transpose $\Gamma$ is (is
  not) positive. Partial transpose of a state $\varrho$ is defined by
  $(\varrho^\Gamma)_{m\mu,n\nu}=\varrho_{m\nu,n\mu}$.}
\bibitem[{PPT()}]{PPTops}
\bibinfo{note}{An operation $\Lambda$ is PPT if $\Gamma \Lambda \Gamma$ is
  completely positive map where $\Gamma$ is partial transpose}.
\bibitem[{con()}]{cont}
\bibinfo{note}{Function $f$ is asymptotically continuous, if for sequences
  $\varrho_n,\sigma_n$ of states on Hilbert space $\varrho_n - \sigma_n \to 0$
  implies ${f(\varrho_n)-f(\sigma_n)\over n }\to 0$.}
\end{thebibliography}
\end{document}